\documentclass[prc,preprint,showpacs,nofootinbib]{revtex4}%
\usepackage{amssymb}
\usepackage{graphicx}
\usepackage{bm}
\usepackage{amsmath}
\usepackage{amsfonts}%
\newcommand{\beq}{\begin{equation}}
\newcommand{\eeq}{\end{equation}}
\newcommand{\bea}{\begin{eqnarray}}
\newcommand{\eea}{\end{eqnarray}}
\newcommand{\ben}{\begin{eqnarray*}}
\newcommand{\een}{\end{eqnarray*}}

\begin{document}
\title{Electron scattering on halo nuclei}
\author{C.A. Bertulani}
\email{bertulani@physics.arizona.edu}
\affiliation{Department of Physics, University of Arizona, 1118 E 4th St., Tucson AZ 85721}

\begin{abstract}
The inelastic scattering of electrons on weakly-bound nuclei is
studied with a simple model based on the long range behavior of the
bound state wavefunction and on the effective-range expansion for
the continuum wavefunctions. Three mechanisms have been considered:
(a) dissociation of halo nuclei by high energy electrons, (b)
dissociation by electrons present in a fixed target, and (c) Coulomb
dissociation. It is shown that the properties of halo nuclei can be
studied in electron-radioactive beam colliders using the
electro-disintegration process. A comparison with fixed-target
experiments is also performed.

\end{abstract}
\maketitle


\draft

\section{\textbf{Introduction}}

A study of properties of weakly-bound neutron-rich, or halo nuclei,
has been carried out intensively worldwide during the last decades
\cite{CS00}. Because of their short beta-decay lifetimes, halo
nuclei are often studied in fragmentation facilities, where they are
produced in-flight. The probes are hadronic, usually stable nuclear
targets. Typically, one uses Coulomb dissociation, stripping,
elastic scattering, etc. \cite{BCH93} as nuclear structure probes.
Such studies are complicated because the reaction mechanisms are not
as well understood as with stable nuclear projectiles. The use of
electromagnetic probes, e.g. electron scattering, is thus highly
desirable. In fact, new experimental facilities for
electron-scattering on unstable nuclear beams are under construction
\cite{GSI}. An accurate determination of charge distributions in
exotic nuclei can be obtained with electrons using inverse
kinematics in a electron-nucleus collider mode \cite{GSI}.
Electronic excitation, or dissociation, of nuclear beams can also be
exploited for a deeper understanding of their structure.

It is the aim of this work to explore basic results of electron
scattering on the simplest of all nuclear halo structures, namely, a
one-neutron halo system. The physics mechanisms and the conditions
for the realization of electron scattering experiments are assessed.
Such study has also an impact in nuclear astrophysics as it allows
to deduce what are the lowest binding energies of halo nuclei
possible in stellar environments, where free electrons are
available.

A high energy beam of weakly-bound neutron-rich nuclei dissociates as it
penetrates a target due to the interaction with the atomic electrons. Since a
heavy element target, e.g., $^{208}$Pb, contains almost 100 electrons,  the
dissociation cross sections are large, assuming that each electron in the atom
scatters independently on the projectile. Moreover, due to the atomic orbital
motion, the innermost electrons have large relative energy with the incoming
nucleus, increasing the dissociation probability. This process is of crucial
importance in designing experiments aiming at studying properties of halo
nuclei with the Coulomb dissociation method.

The dissociation of neutron-rich nuclei, with small neutron
separation energies, in stars can impose stringent limits on the
stellar scenario where these nuclei play a role. For example, if the
r-process proceeds partially out of equilibrium, the neutron
radiative capture cross sections would have to be large enough to
match the electron dissociation cross sections, with the appropriate
neutron and electron density weights.

\section{Electron scattering on neutron halo nuclei}

I will consider the process $e+a\longrightarrow e^{\prime}+b+c$ at small
momentum transfers, $\mathbf{q=}\left(  \mathbf{p}^{\prime}\mathbf{-p}\right)
/\hbar$, such that $qR\ll1$ ($R$ is the nuclear size). For simplicity,
particle $b$ is taken as a neutron and $c$ as a core (inert) nucleus. The
results obtained here are general and can be easily extended to the case of
two-neutron halos.

The differential cross section for this process is given by \cite{EG88}%
\begin{equation}
\frac{d\sigma_{e}}{d\Omega}=\frac{2e^{2}}{\left(  \hbar c\right)  ^{4}}\left(
\frac{p^{\prime}}{p}\right)  \frac{2J_{f}+1}{2J_{i}+1}\frac{EE^{\prime}%
+c^{2}\mathbf{p\cdot p}^{\prime}+m_{e}^{2}c^{4}}{q^{4}}\left\vert \rho
_{fi}\left(  \mathbf{q}\right)  \right\vert ^{2}N_{f}\ , \label{elect_scat}%
\end{equation}
where $E(E^{\prime})$ and $\mathbf{p}(\mathbf{p}^{\prime})$ are the initial
(final) energy and momentum of the electron, respectively. $J_{i}(J_{f})$ is
the initial (final) nuclear spin, and $N_{f}$ is the density of final states
of the nucleus. The nuclear form factor $\rho_{fi}\left(  q\right)  $ is given
by%
\begin{equation}
\rho_{fi}\left(  \mathbf{q}\right)  =\int\rho_{fi}\left(  \mathbf{r}\right)
\ e^{i\mathbf{q\cdot r}}\ d^{3}r\ , \label{form_fact}%
\end{equation}
where $\rho_{fi}\left(  \mathbf{r}\right)  =\psi_{f}^{\ast}\
\psi_{i}$\ is the nuclear charge transition density, with
$\psi_{i}$\ $(\psi_{f})$\ equal to the initial (final) nuclear
wavefunction. The cross section given by eq. \ref{elect_scat} only
includes longitudinal (also called Coulomb) excitations, dominant at
low energy transfers \cite{Sch54,Wei64,Sch66}.

Eq. \ref{elect_scat} is based on the first Born approximation. It
gives good results for light nuclei (e.g. $^{12}$C) and high-energy
electrons. For large-Z nuclei the agreement with experiments is only
of a qualitative nature. The effects of the distortion of the
electron waves have been studied by many authors (see, e.g. ref.
\cite{MF48,Yenn53}). For a rough estimate of this effect, I follow
ref. \cite{Cut67}. For transition densities peaked at the nuclear
surface with radius $R_{0}$, the correction due to Coulomb
distortion
is approximately given by%
\begin{equation}
Q=\frac{d\sigma_{\mathrm{Born}}/d\Omega}{d\sigma_{\mathrm{Corrected}}/d\Omega
}\simeq\frac{1}{1+\beta Ze^{2}/\hbar c},\label{Coulcorr}%
\end{equation}
with%
\begin{align}
\beta &  =\frac{120}{x^{2}}\left\{  -\frac{1}{160x^{3}}\left[  1+\frac{3}%
{2}\cos\left(  2x\right)  +3x\sin\left(  2x\right)  +\frac{x^{2}}{3}\left(
4+5\cos\left(  2x\right)  \right)  +\frac{10}{3}x^{3}\sin\left(  2x\right)
\right]  \right.  \nonumber\\
&  \left.  +\frac{x}{60}\left[  \frac{9}{4}-\cos\left(  2x\right)  \right]
+\frac{x^{2}}{60}\left[  \pi-2\operatorname{Si}\left(  2x\right)  \right]
+n_{1}\left(  2x\right)  \left[  \frac{1}{16x}+\frac{x}{40}\right]  \right\}
,\label{Coulfac}%
\end{align}
where $x=pR_{0}/\hbar$, $\operatorname{Si}$ \ is the sine integral,
$\operatorname{Si}(x)=\int_{0}^{\infty}dt\sin t/t$, and $n_{i}\left(
x\right)  $ is the spherical Bessel function of the second kind. The above
result is valid for monopole ($l=0$) transitions. Corresponding expressions
for higher order transitions are found in ref. \cite{Cut67}.

Table \ref{tab:Coulcorr} shows the correction due to Coulomb
distortion, eq. \ref{Coulcorr}, for $^{11}$Li and $^{19}$C targets
and several electron kinetic energies $K_{e}$. One sees that below
$K_{e}=100$ MeV it is important to account for Coulomb distortion of
the electronic waves. A radius $R_{0}=3.5$ fm was assumed for both
nuclei.  The Coulomb distortion correction decreases approximately
linearly with $R_{0}$.

An additional correction, due to nuclear recoil \cite{EG88}, changes eq.
\ref{elect_scat} by a factor $f_{\mathrm{rec}}\simeq1+(2E_{x}/Mc^{2}%
)\sin(\theta/2)$, where $E_{x}$ is the excitation energy, $M$ is the nuclear
mass, and $\theta$\ is the electron scattering angle. For the dissociation of
weakly-bound nuclei, $E_{x}\ll Mc^{2}$ and this correction is much less
relevant than the distortion of the electronic waves.\ I will neglect the
Coulomb distortion and recoil effects from here on, bearing in mind that they
should be taken into account in a more precise calculation.

\begin{table}[ptbh]
\begin{center}%
\begin{tabular}
[c]{|l|l|l|}\hline
$K_{e}\ $[MeV] & Q$\left(  ^{11}\mathrm{Li}\right)  $ & Q$\left(
^{19}\mathrm{C}\right)  $\\\hline\hline
0.1 & 0.879 & 0.784\\\hline
1 & 0.880 & 0.786\\\hline
10 & 0.887 & 0.797\\\hline
10$^{2}$ & 0.949 & 0.903\\\hline
10$^{3}$ & 0.994 & 0.989\\\hline
\end{tabular}
\end{center}
\caption{The Coulomb correction factor, eq. (\ref{Coulcorr}), for electron
scattering on $^{11}$Li and $^{19}$C and for several kinetic energies, $K_{e}$
(in MeV).}%
\label{tab:Coulcorr}%
\end{table}

In a simplified model for the halo nucleus the radial parts of the initial and
final wavefunctions are represented by single-particle states of the form%
\begin{equation}
u_{i}(r)=A_{i}\ h_{l_{i}}(i\eta r)\ ,\ \ \ \ \ \ \ \ \ \ \ \ u_{f}%
(r)=\cos(\delta_{l_{f}})\ j_{l_{f}}(kr)-\sin(\delta_{l_{f}})\ n_{l_{f}%
}(kr)\label{wav_func}%
\end{equation}
where $\eta$\ is related to the neutron separation energy $S_{n}=\hbar^{2}%
\eta^{2}/2\mu$.$\ h_{l_{i}}(i\eta r)$\ represents the large distance
behavior of the bound state wavefunction, $\mu$\ is the reduced mass
of the neutron + core system and $\hbar k$ their relative momentum
in the final state. $h_{l_{i}}$, $j_{l_{f}}$, and $n_{l_{f}}$ are
the spherical Hankel, Bessel, and Neumann functions, respectively.
$A_{i}\ $\ is the ground state asymptotic normalization coefficient,
which includes the normalization of the neutron single-particle
wavefunction, and a spectroscopic factor \ which accounts for the
many-body aspects. This single-particle picture has been used
previously to study Coulomb excitation of halo nuclei with success
\cite{BB86,BS92,Ots94,MOI95,KB96,TB04}.

The constant $A_{i}$ (with spectroscopic factor equal to the unity)
is used to normalize the bound state wavefunction, and corrects for
the nuclear interaction range, $r_{0}$. In the case of an s-wave
ground state ($l_{i}=0$), one has \cite{HJ87} $A_{i}=\exp\left( \eta
r_{0}\right)  \sqrt{\eta /2\pi\left(  1+\eta r_{0}\right)  }$. For
weakly-bound nuclei, $1/\eta\gg r_{0}$ and
$A_{i}\simeq\sqrt{\eta/2\pi}$. The ground state wavefunction
entering the transition density integral, eq. \ref{form_fact}, is
well represented by the Hankel function $h_{l_{i}}(i\eta r)$. Note
that the wavefunctions in eq. \ref{wav_func} are not orthonormal.
However, the transition density matrix elements of relevance for
Coulomb excitation (and similarly for electron scattering) are
dominated by the outside region ($r>R$) \cite{Ots94,MOI95}. Far from
a resonance, the continuum wavefunction $u_{f}(r)$ is small inside
the nuclear radius. Its asymptotic dependence is well described by
eq. \ref{wav_func}.\

Using eq. \ref{wav_func} the form factor in eq. \ref{form_fact} can
be calculated analytically by expanding $\ e^{i\mathbf{q\cdot r}}$
into multipoles. The results will depend on the parameters $S_{n}$,
$R$, and $\delta_{l_{f}}$. To eliminate the dependence on $R$, the
lower limit of the radial integral in eq. \ref{form_fact} is
extended to $r=0$. The results for an s-wave ground state and the
lowest order continuum angular momenta
($l_{f}=0,1,2$) are particularly simple. They are:%
\begin{align}
\rho_{fi}^{(0)}\left(  q\right)   &  =\frac{e_{\mathrm{eff}}^{(0)}\ \pi A_{i}%
}{qk}\left\{  L-\frac{2k}{-1/a_{0}+r_{0}k^{2}/2}M\right\}  \nonumber\\
\rho_{fi}^{(1)}\left(  q\right)   &  =\frac{e_{\mathrm{eff}}^{(1)}\ i\pi
A_{i}}{q^{2}k^{2}}\left\{  \frac{\eta^{2}+k^{2}+q^{2}}{2}L-2qk-\frac{k^{3}%
}{-1/a_{1}+r_{1}k^{2}/2}\left[  2\eta q+\left(  \eta^{2}+k^{2}+q^{2}\right)
M\right]  \right\}  \nonumber\\
\rho_{fi}^{(2)}\left(  q\right)   &  =\frac{e_{\mathrm{eff}}^{(2)}\ \pi A_{i}%
}{4k^{3}q^{3}}\left\{  8kq\left(  \eta^{2}+k^{2}+q^{2}\right)  -\frac
{3k^{4}+3\left(  \eta^{2}+q^{2}\right)  ^{2}+2k^{2}\left(  3\eta^{2}%
+q^{2}\right)  }{2}\ L\right.  \nonumber\\
&  +\left.  \frac{k^{5}}{-1/a_{2}+r_{2}k^{2}/2}\left[  6\eta\left(  k^{2}%
+\eta^{2}\right)  q+10\eta q^{3}+\left(  3k^{4}+3\left(  \eta^{2}%
+q^{2}\right)  ^{2}+2k^{2}\left(  3\eta^{2}+q^{2}\right)  \right)  M\right]
\right\}  {}{}{},\label{effr}%
\end{align}
where%
\begin{equation}
L=\ln\left(  \frac{\eta^{2}+\left(  k+q\right)  ^{2}}{\eta^{2}+\left(
k-q\right)  ^{2}}\right)  ,\ \ \ \ \ \ \ \ \text{and}\ \ \ M=\tan^{-1}\left(
\frac{k-q}{\eta}\right)  -\tan^{-1}\left(  \frac{k+q}{\eta}\right)  \ .
\end{equation}
In these equations $e_{\mathrm{eff}}^{(\lambda)}$ $=eZ\left(  -1/A\right)
^{\lambda}$ is the neutron-core effective charge which depends on the
transition multipolarity $\lambda$ ($\lambda=l_{f}$ for $l_{i}=0$). \ The
effective range approximation $k^{2l+1}\cot\delta_{l}=-1/a_{l}+r_{l}k^{2}/2$
has been used, where the parameters $a_{l}$ and $r_{l}$ are the scattering
length and the effective range, respectively. Notice that only for $l=0$ the
scattering length and effective range have dimensions of length.

The $l=0$ form factor has a large sensitivity to the orthogonality
of the wavefunctions. If one assumes a zero-range potential for the
neutron-core interaction, the scattering wavefunction, orthogonal to
the bound-state wavefunction, is given by
$\psi_{\mathbf{k}}^{(+)}(\mathbf{r})=\exp (i\mathbf{k\cdot
r})-\exp(ikr)/[(\eta+ik)r]$. The s-wave scattering length is then
just $a_{0}=1/\eta$. Using this value, together with $r_{0}=0$, in
the equation for $\rho_{fi}^{(0)}\left(  q\right)  $ leads to a
large cancelation between the first and second terms. The $l=1,2$
form factors are also very sensitive to the scattering lengths and
effective ranges. For example using $a_{1}\simeq5$ fm$^{3}$ and
$r_{1}=0$ fm$^{-1}$ reduces the magnitude of $\rho_{fi}^{(1)}\left(
q\right)  $ by 10\% for scattering at forward angles. These results
show that it is very important to include the correct energy
dependence of the phase-shifts to obtain an accurate description of
electron scattering off halo nuclei \cite{TB04}.

In what follows, I will use $e_{\mathrm{eff}}^{(\lambda)}=e,$ $R=0$,
$\mu=m_{N}$ (nucleon mass), and neglect the terms containing the
effective-range expansion parameters in eq. \ref{effr}. These
approximations are not necessary but, with these choices, the
numerical results will not depend on the charges and mass parameters
of a particular nucleus; only on its neutron separation energy
$S_{n}$.

\begin{figure}[t]
\begin{center}
\includegraphics[
height=3.2906in,
width=3.3261in
]{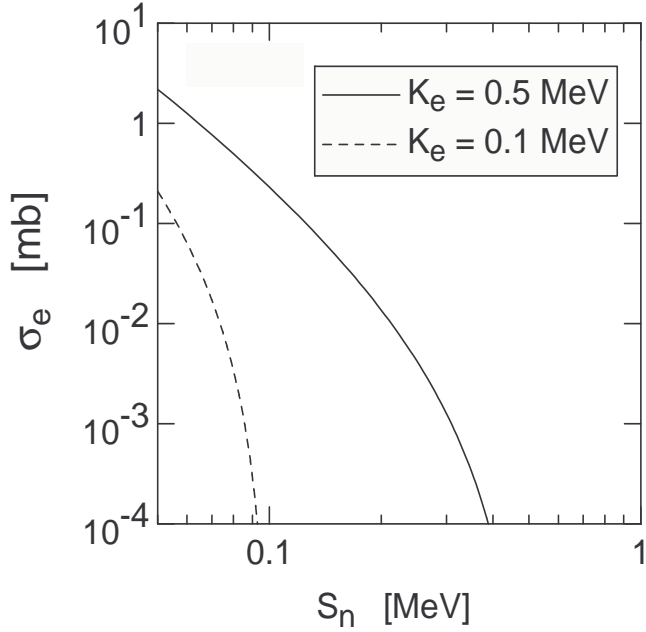}
\end{center}
\caption{Cross sections for electron induced breakup as a function of the
separation energy, $S_{n}$, and for electron bombarding energies equal to 0.1
(dashed) and 0.5 MeV (solid).}%
\label{sn}%
\end{figure}

The total electron-disintegration cross section is obtained from eq.
\ref{elect_scat}, with the density of states given by
$N_{f}=d^{3}k/\left( 2\pi\right)  ^{3}$, and integrating over
$\mathbf{k}$ and $\Omega.$ Figure \ref{sn} shows the
electro-dissociation cross sections obtained by a numerical
integration of eq. \ref{elect_scat}, as a function of the separation
energy, $S_{n}$, for electron bombarding energies equal to 0.1 and
0.5 MeV, respectively. One observes that only for very low neutron
separation energies ($S_{n}\lesssim50$ keV) the
electro-disintegration cross section becomes larger than 1 mb. If a
more realistic model for $l=1$ transitions is used, the
cross section will be further reduced by: (a) a factor $\left(  Z_{a}%
/A_{a}\right)  ^{2}\lesssim1/4$ due to the effective charge, (b) by
properly orthogonalized wavefunctions, and (c) by the energy
dependence of the phase-shifts. 

Figure \ref{sn} also shows that the electron dissociation cross section
increases appreciably with the electron energy. It is thus instructive to
study the dependence of the cross section on the electron energy at high
energies. The electron energy will be considered to be much larger than the
energy transfer in the dissociation, i.e. $E\gg\Delta E=E_{x}$ ($E_{x}$
denotes the excitation energy). The scattering is peaked at forward angles
and, from kinematics, $q=k^{\prime}\cos\theta-k\simeq\Delta k\simeq
E_{x}/\hbar c$. For energy transfers $E_{x}$ \ of the order of a few MeV, one
also has $q\ll p\mathbf{,\ }\eta$. Using eqs. \ref{effr} one obtains for the
leading multipolarity ($l=1$)%
\begin{equation}
\rho_{fi}^{(1)}\left(  q\right)  \cong\frac{4\pi i\ e\ A\ qk}{\left(
k^{2}+\eta^{2}\right)  ^{2}}\ .\label{rhoapp}%
\end{equation}

Using eqs. \ref{rhoapp} and \ref{elect_scat} one obtains%
\begin{equation}
\frac{d\sigma_{e}}{d\Omega dE_{x}}=\frac{48\sqrt{2}}{\pi}\frac{e^{2}\left[
e_{\mathrm{eff}}^{(1)}\right]  ^{2}p^{2}}{\hbar^{2}\mu c^{2}}\frac{1}{q^{2}%
}\frac{\sqrt{S_{n}}\left(  E_{x}-S_{n}\right)  ^{3/2}}{E_{x}^{4}%
}.\label{sigeapp}%
\end{equation}
The solid scattering angle can be related to the momentum transfer
by means of $d\Omega\cong2\pi\hbar^{2}qdq/p^{2}.$ The minimum
momentum transfer for an excitation energy $E_{x}$ is given by
$q_{\min}=\Delta k\cong
E_{x}/\hbar c$, so that the integral over the scattering angle yields%
\begin{equation}
\frac{d\sigma_{e}}{dE_{x}}=96\sqrt{2}\ \frac{e^{2}\left[  e_{\mathrm{eff}%
}^{(1)}\right]  ^{2}}{\mu c^{2}}\ \frac{\sqrt{S_{n}}\left(  E_{x}%
-S_{n}\right)  ^{3/2}}{E_{x}^{4}}\ \ln\left(  \frac{pc}{E_{x}}\right)
.\label{sigex}%
\end{equation}
Eq. \ref{sigex} shows that, for large $p$, the energy spectrum in
electro-disintegration depends weakly on the electron energy through
the logarithm function. This means that there is no great advantage
(in terms of
number of events) in increasing the electron energy when $E_{e}\gg m_{e}c^{2}%
$. From eq. \ref{sigex} one also sees that the energy spectrum
increases sharply starting at $E_{x}=S_{n}$, peaks at
$E_{x}=8S_{n}/5$, and decreases with $E_{x}^{-5/2}$ at large
energies. This is the same characteristic spectrum as found in
Coulomb dissociation of halo nuclei \cite{BS92}.

The integral over the excitation energy gives, to leading order,%
\begin{equation}
\sigma_{e}\left(  p\right)  =6\pi\sqrt{2}e^{2}\left[  e_{\mathrm{eff}}%
^{(1)}\right]  ^{2}\frac{1}{\mu c^{2}S_{n}}\ln\left(  \frac{pc}{S_{n}}\right)
. \label{sigef}%
\end{equation}

For stable nuclei, with $S_{n}\simeq$ few MeV, the
electron-disintegration cross section is small. The dependence of
eq. \ref{sigef} on the inverse of the separation energy is most
important for loosely bound nuclei. Using $S_{n}=100$ keV,
$E_{e}\cong pc=10$ MeV, $e_{\mathrm{eff}}^{(1)}=e,$ and $\mu
c^{2}=10^{3}$ MeV, equation \ref{sigef}\ yields 25 mb for the
dissociation cross section by high energy electrons. Note that the
above equations are valid only if $E_{e}\gg m_{e}c^{2}$. They show
that the electro-disintegration cross section increases very slowly
with the electron energy. In contrast, as shown in figure \ref{sn},
at low electron energies the cross sections increase much faster
with $E_{e}$.

The arguments used here are only valid for dissociation (breakup)
experiments. In the case of electron excitation of bound states, the
matrix elements can become large for small excitation energies and
cases where there is a large overlap of the wavefunctions.
Consequently, the cross section can be much higher when these
conditions are met.

\section{Dissociation of halo nuclei beams on a fixed target}

\subsection{Dissociation by atomic electrons in the target}

I use the Thomas-Fermi model to describe the electronic distribution in an
atom. This approximation is well known, being described in many textbooks
(see, e.g., ref. \cite{Fri90}). In this model, the electron density as a
function of the distance from the atomic nucleus with charge $Ze$ is given by%
\begin{equation}
\rho\left(  r\right)  =\frac{1}{3\pi^{2}}\left[  2\frac{m_{e}}{\hbar^{2}%
}Ze^{2}\frac{\Phi\left(  x\right)  }{r}\right]  ^{3/2},\ \ \mathrm{where}%
\ \ \ x=br,\ \ \ \mathrm{and}\ \ \ b=2\left(  \frac{4}{3\pi}\right)
^{2/3}\frac{m_{e}}{\hbar^{2}}e^{2}Z^{1/3}.
\end{equation}
\ The function $\Phi\left(  x\right)  $ is the solution of the Thomas-Fermi
equation%
\begin{equation}
\frac{d^{2}\Phi}{dx^{2}}=\frac{\Phi^{3/2}}{x^{1/2}}\ .
\end{equation}
Numerical solutions of this equation date back to refs. \cite{Bak30,BC31}. An
excellent approximation was found by Tietz \cite{Tie56}:%
\begin{equation}
\Phi\left(  x\right)  =\frac{1}{\left(  1+ax\right)  ^{2}}%
,\ \ \ \mathrm{where}\ \ \ \ \ \ \ \ a=0.53625.
\end{equation}

The probability density (normalized to $Z$) to find an electron with momentum
$p$ is given by%
\begin{equation}
\mathcal{P}(p)=\left\vert D\left(  \mathbf{p}\right)  \right\vert
^{2},\ \ \ \ \mathrm{where}\ \ \ D\left(  \mathbf{p}\right)  =\frac{1}{\left(
2\pi\right)  ^{3/2}}\int d^{3}re^{i\mathbf{p\cdot r}}\ \sqrt{\rho\left(
r\right)  }.\label{FT1}%
\end{equation}
The electronic density $\rho\left(  r\right)  $\ has to be Lorentz transformed
to the frame of reference of the projectile nucleus. Assuming a straight-line
projectile motion with impact parameter $b$ from the atomic center, the
transformed density is%
\begin{equation}
\rho^{\prime}\left(  r\right)  =\gamma\rho\left(  \sqrt{b^{2}+\gamma^{2}z^{2}%
}\right)  ,\label{Dboost}%
\end{equation}
where $\gamma=\left(  1-v^{2}/c^{2}\right)  ^{-1/2}$ is the Lorentz factor,
and $v$ is the projectile velocity.

The Fourier transform in eq. \ref{FT1}\ becomes%
\begin{equation}
D^{\prime}\left(  \mathbf{p}\right)  =\frac{1}{\left(  2\pi\right)  ^{3/2}%
}\int d^{3}r\ e^{i\mathbf{p\cdot r}}\ \sqrt{\rho^{\prime}\left(  r\right)
}=\frac{1}{\left(  2\pi\right)  ^{3/2}}\frac{1}{\sqrt{\gamma}}\int
d^{3}r^{\prime}\ e^{i\mathbf{P\cdot r}^{\prime}}\ \sqrt{\rho\left(  r^{\prime
}\right)  }\ ,\label{FT2}%
\end{equation}
where
\begin{equation}
\mathbf{P=}\left(  \mathbf{p}_{t},p_{z}/\gamma\right)
,\ \ \ \ \ \ \ r^{\prime}=\left(  \mathbf{b},\gamma z\right)  \ ,
\end{equation}
with $\mathbf{p}_{t}\left(  p_{z}\right)  $ being the transverse
(longitudinal) momentum.

Since $\rho\left(  r\right)  $\ is spherically symmetric, eq. \ref{FT2}\ can
be rewritten as%
\begin{equation}
D^{\prime}\left(  \mathbf{p}\right)  =\sqrt{\frac{2}{\pi\gamma}}\frac{1}%
{P}\int dr\ r\ \sin\left(  Pr\right)  \ \sqrt{\rho\left(  r\right)  }\ .
\end{equation}

\begin{figure}[t]
\begin{center}
\includegraphics[
height=2.9689in, width=3.2379in ]{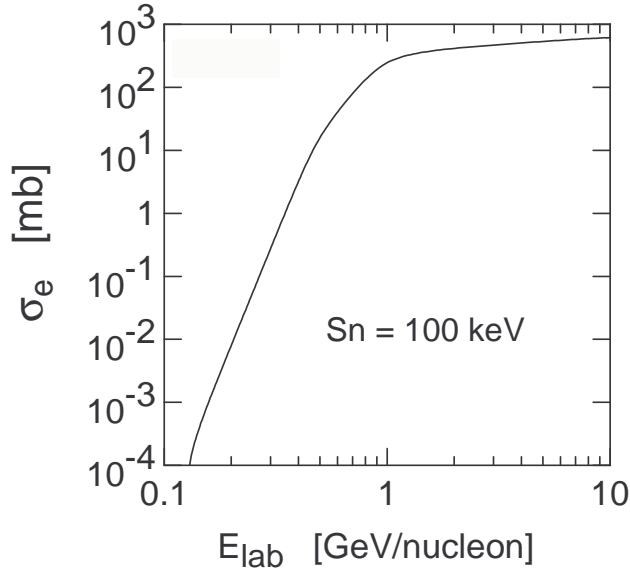}
\end{center}
\caption{Cross section for the electro-dissociation of a neutron-halo nucleus
impinging on a lead target, as a function of the bombarding energy.  A
separation energy equal to 100 keV was used.}%
\label{ate}%
\end{figure}

For an atom at rest, very few electrons have orbital kinetic
energies larger than 100 keV. In the case of $^{92}$U only 3\% of
the electrons (2 electrons!)\ have kinetic energies larger than
that. But in the reference frame of a 100 MeV/nucleon projectile,
50\% of the electrons have energies greater than 100 keV.

Assuming that each electron scatters independently, the total dissociation
cross section by the target atomic electrons is given by%
\begin{equation}
\sigma_{e}^{\left(  TAE\right)  }\left(  p\right)  =\int d^{3}p\ \mathcal{P}%
^{\prime}(\mathbf{p})\ \sigma_{e}\left(  p\right)  =2\pi\int_{0}^{\infty
}dp_{t}\ p_{t}\int_{-\infty}^{\infty}dp_{z}\ \mathcal{P}^{\prime}(p_{z}%
,p_{t})\ \sigma_{e}\left(  p\right)  \ .
\end{equation}
The separation of the above integral into longitudinal and transverse momenta
is convenient because only the longitudinal momentum component of the
electrons is relevant for the dissociation of the projectile.

Figure \ref{ate} shows the dissociation cross section for a halo nucleus, with
separation energy $S_{n}=100$ keV, incident on a Pb target as a function of
the bombarding energy. Although the cross sections are small for incident
energies equal to a few hundred MeV/nucleon, they increase drastically as the
bombarding energy becomes close to 1 GeV/nucleon. At 10 GeV/nucleon the
dissociation cross section is of the order of 1 barn.

Comparing the above results with those obtained in section 2, we
notice that there are different energy scales for
electro-disintegration on fixed targets (by atomic electrons) and on
a collider-beam mode. This is due to the Lorentz transformation and
to the large density (compared to an electron beam) of electrons in
a heavy atom. Thus, with beams of halo nuclei  with a few
GeV/nucleon one could, in principle, perform similar studies as with
electron-radioactive beam colliders. The disadvantage is that the
Coulomb dissociation cross sections of loosely-bound nuclei are much
larger, as shown in the next section.

\subsection{Coulomb dissociation}

Coulomb dissociation of halo nuclei has been considered long time ago
\cite{BB86}. For the leading electric dipole transitions from an s- to a
p-wave, the Coulomb dissociation cross section is given by \cite{BB86,BS92}%
\begin{equation}
\frac{d\sigma_{C}}{dE_{x}}=\frac{32}{3}\frac{Z^{2}e^{2}\left[  e_{\mathrm{eff}%
}^{(1)}\right]  ^{2}}{\hbar^{2}c^{2}}\ \frac{\sqrt{S_{n}}\left(  E_{x}%
-S_{n}\right)  ^{3/2}}{E_{x}^{4}}\ \ln\left(  \frac{\gamma\hbar c}{\delta
E_{x}R}\right)  ,\label{sigcoul}%
\end{equation}
where $Z$ is the nuclear target charge, $\delta=0.681...$, and $R$
is the strong interaction radius ($R\simeq R_{P}+R_{T}$). Note the
similarity  with eq. \ref{sigex} in the dependence on the excitation
energy $E_{x}$, because the dipole operator is the same in both
cases. However, the argument of the logarithm is different because
of the small electron mass. Moreover, the coherent electric field of
the projectile yields a factor $Z^{2}$\ which substantially
increases the Coulomb dissociation cross section for large-$Z$
targets.

The total cross section for Coulomb dissociation as a function of the
bombarding energy (i.e., as a function of $\gamma$) is given by%
\begin{equation}
\sigma_{C}=\frac{2\pi}{3}\frac{Z^{2}e^{2}\left[  e_{\mathrm{eff}}%
^{(1)}\right]  ^{2}}{\hbar^{2}c^{2}}\ \frac{\hbar^{2}}{\mu S_{n}}\ \ln\left(
\frac{\gamma\hbar c}{\delta S_{n}R}\right)  .
\end{equation}

Using the same values listed after eq. \ref{sigef}, for 10
GeV/nucleon projectiles impinging on Pb targets, yields cross
sections of approximately 24 barns. This is much larger than that
due to the dissociation by electrons in the target. But the
contribution of the later process comprises 5\% of the total
disintegration cross section, and should be considered in
experimental analysis.

\section{Conclusions}

In this article the inelastic scattering of electrons off halo
nuclei was studied, with emphasis on the energy dependence of the
dissociation cross sections. It is shown that the cross sections for
electro-dissociation of weakly-bound nuclei reach ten milibarns for
10 MeV electrons and increase logarithmically at higher energies.
This means that extracting information about the continuum structure
of weakly-bound nuclei (e.g. scattering lengths and effective
ranges, as in eq. \ref{effr}) can only be done if the intensity of
the radioactive beam is very large, or if the collider allows for a
large number of sequential interactions between the electrons and
the nuclei at different crossing points. This conclusion can be
drawn from figure \ref{sn}, where a steep decrease of the
dissociation cross section with $S_{n}$ is seen. Halo breakup
experiments (common in fixed-target radioactive beam facilities) are
difficult to carry out in electron-radioactive beam colliders, but
not impossible if $S_{n}$ is small.

A new facility is under construction at the GSI/Darmstadt, Germany.
Experiments in a collider mode are planned so that electron beams
will cross radioactive beams with center-of-mass energies of 1.5
GeV, i.e. 0.5 GeV electrons impinging on a 740 MeV/A counter
propagating ion \cite{GSI,Haik}. For light, neutron-rich, nuclei
luminosities of $10^{29}$/cm$^{2}$.s are expected. The approximate
eq. \ref{sigef} yields cross sections of the order of 1 mb for
$S_{n}\simeq1$ MeV, what means an estimated 100 events/second.

I have also shown that electrons present in a fixed nuclear target
access similar scattering conditions as in an electron-radioactive
beam collider. However, Coulomb excitation cross sections are much
larger in the case of a heavy nuclear target. In view of the
scientific impact of an electron-radioactive beam facility these
results are useful for guidance in planning future experiments. The
role of electron (and photon) scattering on exotic nuclei in stellar
environments is also of interest for stellar modeling and work in
this direction is in progress.

\section{Acknowledgements}

I thank useful discussions with P.G. Hansen, H. Schatz, H. Simon, U. van Kolck
and V. Zelevinsky. This research was supported in part by the Department of
Energy under Grant No. DE-FG02-04ER41338.


\begin{thebibliography}{99}                                                                                               %


\bibitem {CS00}R.F. Casten and B.M. Sherrill, \textit{Prog. Part. Nucl. Phys.}
\textbf{45} (2000) S171.

\bibitem {BCH93}C.A. Bertulani, L.F. Canto and M.S. Hussein, Phys. Rep.
\textbf{226}, 281 (1993).

\bibitem {GSI}FAIR: Facility for Antiproton and Ion Research, Conceptual
Design Report, GSI, 2002, p. 162.

\bibitem {EG88}J.M. Eisenberg and W. Greiner, \textquotedblleft Excitation
Mechanisms of the Nucleus", (North-Holland, Amsterdam, 1988).

\bibitem {Sch54}L.I. Schiff, Phys. Rev. \textbf{96}, 765 (1954).

\bibitem {Wei64}L.J. Weigert and J.M. Eisenberg, Nucl. Phys. \textbf{53}, 508 (1964).

\bibitem {Sch66}F. Scheck, Nucl. Phys. \textbf{77}, 577 (1966).

\bibitem {MF48}W.A. McKinley and H. Feshbach, \ Phys. rev. \textbf{74}, 1759 (1948).

\bibitem {Yenn53}D.R. Yennie, D.G. Ravenhall, and R.R. Wilson, Phys. Rev.
\textbf{92}, 1325 (1953); Phys. Rev. \textbf{95}, 500 (1954).

\bibitem {Cut67}L.S. Cutler, Phys. Rev. \textbf{157}, 885 (1967).

\bibitem {BB86}C.A. Bertulani and G. Baur, Nucl. Phys. \textbf{A480}, 615
(1988). {\small Note that a factor 1/3 is missing in eqs. 3.2b and 4.3b of
this reference}.

\bibitem {BS92}C.A. Bertulani and A. Sustich, Phys. Rev. \textbf{C 46}, 2340 (1992).

\bibitem {Ots94}T. Otsuka et al., Phys. Rev. \textbf{C 49}, R2289 (1994).

\bibitem {MOI95}A. Mengoni, T. Otsuka and M. Ishihara, Phys. Rev. C
\textbf{52}, R2334 (1995).

\bibitem {KB96}D.M. Kalassa and G. Baur, J. Phys. \textbf{G 22}, 115 (1996).

\bibitem {TB04}S. Typel and G. Baur, Phys. Rev. Lett. \textbf{93}, 142502 (2004).

\bibitem {HJ87}P.G. Hansen and B. Jonson, Europhys. Lett. \textbf{4}, 409 (1987).

\bibitem {Fri90}H. Friedrich, \textquotedblleft Theoretical Atomic Physics",
(Springer-Verlag, Heidelberg, 1990).

\bibitem {Bak30}E.B. Baker, Phys. Rev. \textbf{36}, 630 (1930).

\bibitem {BC31}V. Bush and S.H. Caldwell, Phys. Rev. C \textbf{38}, 1898 (1931).

\bibitem {Tie56}T. Tietz, J. Chem. Physics \textbf{25}, 787 (1956); Z.
Naturforsch. \textbf{23a}, 191 (1968).

\bibitem {Ho58}R. Hofstadter, Rev. Mod. Phys. \textbf{28}, 214 (1956).

\bibitem {Ra58}D. G. Ravenhall, Rev. Mod. Phys. \textbf{30}, 430 (1958).

\bibitem {Haik}Haik Simon, private communication.
\end{thebibliography}
\end{document}